\documentclass[12pt]{article}
\usepackage{graphicx,amsmath,amssymb,amsthm,euscript}
\newcommand{\ssy}[5]{#1  \emph{#2}\ {\bf #3} (#4) #5\rlap{.}}
\newcommand{\karti}[4]{\begin{figure}[#1]\begin{center}
\includegraphics[height=#2]{#3}
\end{center}\caption{#4}\end{figure}}
\newenvironment{sv-va}{\begin{enumerate}

        }
{
        \end{enumerate}}
\newcommand*{\ogr}[2]{#1\, \rule[-0.76em]{0.4pt}{1.4em} \,
\raisebox{-.55em}{$\displaystyle{}_{#2}$}  }
\newcounter{rem}
\newcommand{\rem}{\refstepcounter{rem}\par\noindent\emph{Remark \therem.} }
\renewcommand{\leq}{\leqslant}
\renewcommand{\geq}{\geqslant}%
\newcommand{\rmd}{\mathrm{d}}
\newcommand*{\hence}{\Longrightarrow}
\newcommand{\tu}{\ensuremath{\mathcal T}}
\newcommand{\mwh}{ m}
\newcommand{\hor}{\ensuremath{h}}
\newcommand{\s}{\ensuremath{\EuScript S}}
\newcommand{\os}{$M_\text{wh}$}
\newcommand*{\oh}[1]{{\hat{#1}}}
\newcommand*{\Un}[1]{{\mathring{#1}}}
\title{Evaporation induced traversability of the Einstein--Rosen
wormhole.}
\author{S. Krasnikov\\
\emph{The Central Astronomical Observatory at Pulkovo}}
\date{}
\begin{document}
\maketitle
\begin{abstract}
Suppose, the Universe comes into existence (as classical spacetime)
already with an empty spherically symmetric macroscopic wormhole present
in it. Classically the wormhole would evolve into a part of the
Schwarzschild space and thus would not allow any signal to traverse it. I
consider semiclassical corrections to that picture and build a model of
an evaporating wormhole. The model is based on the assumption that the
vacuum polarization and its backreaction on the geometry of the wormhole
are weak. The lack of information about the era preceding the emergence
of the wormhole results in appearance of three parameters which --- along
with the initial mass --- determine the evolution of the wormhole. For
some values of these parameters the wormhole turns out to be long-lived
enough to be traversed  and to transform into a time machine.
\end{abstract}
\section{Introduction}
The question as to whether there are  traversable wormholes in the
Universe is at present among the most important problems of classical
relativity. The reason is that in the course of its evolution a spacetime
with such a wormhole is apt to develop the Cauchy horizon \cite{Tho}. At
one time it was  believed that closed timelike curves must lurk beyond
the horizon and it was commonplace to tie existence of wormholes with
possibility of time machines. Later it has become clear that the two
phenomena are not directly connected --- the spacetime \emph{always} can
be extended through the Cauchy horizon in infinitely many ways, all these
extensions being equal (since \emph{none} of them is globally
hyperbolic), and \emph{always} some of them are causal
\cite{eotm}. Nevertheless, the fact remains that having a wormhole one
can (try to) force the spacetime to choose between a number of
continuations and we have no idea as to the criteria of the choice%
\footnote{The Cauchy horizons are also expected inside the black holes,
but we are  protected from whatever is beyond them by the event horizons.
At the same time a wormhole enables a mad scientist with finite resources
to destroy the universe, as is romantically put by Wald
\cite{WCCS}. For discussion on quantum effects that may, or may not save
the Universe see
\cite{qu}.}. That is the existence of traversable wormholes would possibly
imply the existence of an unknown classical (though non-local!) law
governing the evolution of the Universe.

The process of emergence of the classical spacetime from what precedes it
is not clear yet (to say the least). So, it is well possible that the
whole problem is spurious and there are no traversable wormholes just
because they have never appeared in the first place. In principle, one
can speculate that there is a mechanism suppressing formation of any
topological `irregularity' at the onset of the classical universe.
However, at present nothing suggests the existence of  such a mechanism
and it seems reasonable to pose the question: assuming a wormhole did
appear in the end of the Planck era, what would happen with it? Would it
last for long enough to threaten global hyperbolicity?

Traditionally in searching for traversable wormholes one picks a
stationary (and hence traversable) wormhole and looks for matter that
could support it. However, in none of the hitherto examined wormholes
 the required matter looks too realistic. In some cases it is
phantom matter with a prescribed equation of state \cite{sush_phantom},
in some others
--- classical scalar field \cite{Bar_Vis}. True, two wormholes are known
\cite{HPS,MWH} the matter content of which is less exotic in that it at least  obeys
the weak energy condition (all necessary \cite{Tho,FSW} violations of the
latter being provided by the vacuum polarization). However, the first of
them  has the throat $67l_\text{Pl}$ wide and therefore, being nominally
a wormhole, can scarcely be called traversable. The second is macroscopic
(arbitrarily large, in fact), but needs some classical matter. Though
this matter does satisfy the weak energy condition (WEC), nothing at the
moment is known about how realistic it is in  other respects. In this
paper I take a different approach: first, I fix the initial form and the
matter content of the wormhole trying to choose them as simple as
possible (the hope is that the simpler is the model the better are the
chances that it reflects general properties of the real wormholes). Then
I subject it to the (semiclassical) Einstein equations
\begin{equation*}
G_{\mu\nu} = 8\pi T^\text{c}_{\mu\nu}+8\pi T_{\mu\nu}
\end{equation*}
(here $T_{\mu\nu}$ is the expectation value of the quantum stress-energy
tensor and $T^\text{c}_{\mu\nu}$ is the contribution of the classical
matter) and trace the evolution of this presumably realistic wormhole
testing it for traversability.

The wormhole under consideration  --- I shall denote it \os\ --- comes
into being in the end of the Planck era as  the Schwarzschild space with
 mass $m_0$ (to be more precise, as a three-dimen\-sional subspace \s\
thereof), hence the name Einstein--Rosen wormhole. The form of \s\ is
defined by trans-Planckian physics that gives birth to the wormhole. I
set three conditions on \s, of which only one seems to lead to noticeable
loss in generality. Each of the allowed \s\ is characterized by three
numbers
--- $\kappa_R$, $\kappa_L$, and $\varpi$. For a given mass $\varpi$ is
related to the minimal possible radius of
\s\ and, when $\varpi$ is fixed, $\kappa_{R(L)}$ loosely speaking measures
the delay between the end of the Planck era near the throat and in the
remote parts of the right (left) asymptotically flat region (I mostly
consider an `inter-universe wormhole'
\cite{viss}, i.~e.\ a spacetime with two  asymptotically flat
regions connected by a throat; an `intra-universe wormhole' is
constructed in section~\ref{sec:conc} by identifying parts of these
regions, correspondingly a new parameter $d$ --- the distance between the
mouths --- appears).

The wormhole  is taken to be empty: $T^\text{c}_{\mu\nu}=0$ (for reasons
of simplicity again). Hence, classically it would be just (a part of) the
Schwarzschild space $M_\text{S}$, which is a standard of
non-traversability \cite{Tho}. But the Schwarzschild black hole,  as is
well known, evaporates, that is quantum effects in $M_\text{S}$ give rise
to a non-zero vacuum stress-energy tensor $\mathring T_{\mu\nu}$. So, by
the Einstein equations \os\ is anything but $M_\text{S}$. Determination
of its real geometry
is, in fact,  a longstanding problem, see e.~g.\ \cite{obz} for
references and \cite{HK} for some discussion on its possible relation to
the wormholes. In this paper I make no attempts to solve it. It turns out
that to study traversability of a wormhole all one needs to know is the
metric in the immediate vicinity of the apparent horizons and,
fortunately, for  wormholes with the proper values of $\varpi$ this
--- simpler --- problem can be solved separately.

To that end I make a few assumptions based on the idea that quantum
effects are relatively weak. Roughly, I assume that the system (Einstein
equations + quantum field equations) has a solution \os\ with the
geometry resembling that of the Schwarzschild space --- and coinciding
with the latter on \s\ --- and with $T_{\mu\nu}$ close to that of the
conformal scalar field in the Unruh vacuum (what exactly the words
`resembling' and `close' mean in this context is specified in
section~\ref{sec:WE}). Though the above-mentioned assumptions are quite
usual and on the face of it seem rather innocent, in \emph{some}
situations, as we shall see, they cannot be true (which on the second
thought is not surprising --- one does not expect the vacuum polarization
to be weak near the singularity, or in the throat at the moment of its
maximal expansion). Therefore the consideration will be restricted to the
class of wormholes with~$\varpi\in (1,\frac{\sqrt 5 +1}{2})$.

Throughout the paper the Planck units are used: $G=c=\hbar=1$ and the
mass $m_0$ is supposed to be large in these units.

\section{The model and the assumptions}

\subsection{The Schwarzschild spacetime}
We begin with recapitulating some facts about the Schwarzschild space,
which will be needed later.

The eternal (though non-static) spherically symmetric empty wormhole is
described by the Schwarzschild  metric, which we shall write in the form
 \begin{equation}\label{eq:Sch}
  \rmd s^2= -\mathring{F} ^2({u },{v })\rmd {u }\rmd {v } +
  \mathring{r}^2({u },{v })(\rmd\theta^2 + \cos^2\theta\,\rmd\phi),
\end{equation}
where
\begin{equation}\label{eq:strSch}
  \mathring{F} ^2=16m_0 ^2x^{-1}e^{-x},\qquad
\mathring{r}= 2m_0  x
\end{equation}
and the function $x({u },{v })$ is defined implicitly by the equation
\begin{equation}\label{eq:strSch'}
 {u }{v }=(1-x)e^x.
\end{equation}
It is easy to check that the following relations hold
\begin{subequations}
\begin{equation}\label{eq:rvs}
  \mathring{r},_{v }=-\frac{{2m_0  u }}{xe^x}
\end{equation}
\begin{equation}\label{eq:rus}
  \mathring{r},_{u }=-\frac{{2m_0  v }}{xe^x}= 2m_0  \frac{x-1}{u x}
\end{equation}
\begin{equation}\label{eq:xuv}
\mathring r,_{{u }{v }}=-2m_0 \frac{e^{-x}}{x^3}
\end{equation}
\begin{equation}\label{eq:phiu}
 \mathring\varphi ,_{u }=-\tfrac{1}{2}(\ln x +x),_{u }=
 -\frac{1+x}{2x}x,_{u }
  \qquad\text{where}\quad
 \mathring\varphi \equiv\ln \mathring{F}.
\end{equation}
\end{subequations}

In the Unruh vacuum the expectation value of the stress-energy tensor of
the conformal scalar field has the following structure:
\begin{align*}
  4\pi \Un{T}_{v v }&=\tau_1 \mathring{r},_{u }^{-2}
\\
  4\pi \Un{T}_{u u }&=\tau_2 \mathring{r},_{u }^2 m_0 ^{-4}
\\
  4\pi \Un{T}_{u v }&=\tau_3  m_0 ^{-2}
\\ \label{eq:tthth}
4\pi \Un{T}_{\theta\theta}&=4\pi\cos^{-2}\theta\,\Un{T}_{\phi\phi}
=\tau_4 m_0^{-2}
\end{align*}
(all remaining components of $\Un{T}_{\mu\nu}$ are zero due to the
spherical symmetry), where $\tau_i$ are  functions of $x$, but not of ${u
},{v }$, or $m_0 $ separately.
 What is known about $\tau_i(x)$ supports the idea that in the Planck
units they are small. In particular, $|\tau_i(1)|\lesssim 10^{-3} $ and
$K$ defined in \eqref{eq:defK} is $\approx 5\cdot 10^{-6}$ as follows
from the results of
\cite{ChrFu} and \cite{Elster}, see Appendix~\ref{app:1}. At the horizons
$\mathring{h}$, which in this case are the surfaces $x=1$,
\begin{subequations}\label{eq:TonSchor_S}
\begin{equation}\label{eq:defK}
 \mathring{r},_{u }^{2}\ogr{  \Un{T}_{v v }}{\mathring{h}}=
 \frac{\tau_1(1)}{4\pi}=
 -\frac{\mathring{F}^4(1)K}{16m_0 ^4},
 \qquad K\equiv - \frac{\tau_1(1)e^2}{64\pi} ,
\end{equation}
\begin{equation}
  \mathring{r},_{u }^{-2} \Un{T}_{u u }\ogr{}{\mathring{h}}=
  \frac{\tau_2 (1)}{4 \pi m_0^4} ,
\end{equation}
and
\begin{equation}
 \ogr{  |\Un{T}_{u v }|}{\mathring{h}}=
 \frac{\tau_3(1)}{4\pi m_0^2}\ll
 \frac{\mathring{F}^2}{64\pi m_0 ^2} .
\end{equation}
\end{subequations}

\subsection{The geometry of the  Einstein--Rosen wormhole}\label{sec:monot}

The wormhole \os\ being discussed is a spacetime with the metric
\begin{equation}\label{eq:ourmetr}
  \rmd s^2= -F^2({u},{v})\,\rmd {u}\rmd {v} + r^2({u},{v})(\rmd\theta^2 +
  \cos^2\theta\,\rmd\phi).
\end{equation}
To express the idea that the wormhole is `initially Schwarzschildian' we
require that there should be a  surface \s\
\karti{t,b}{0.5\textwidth}{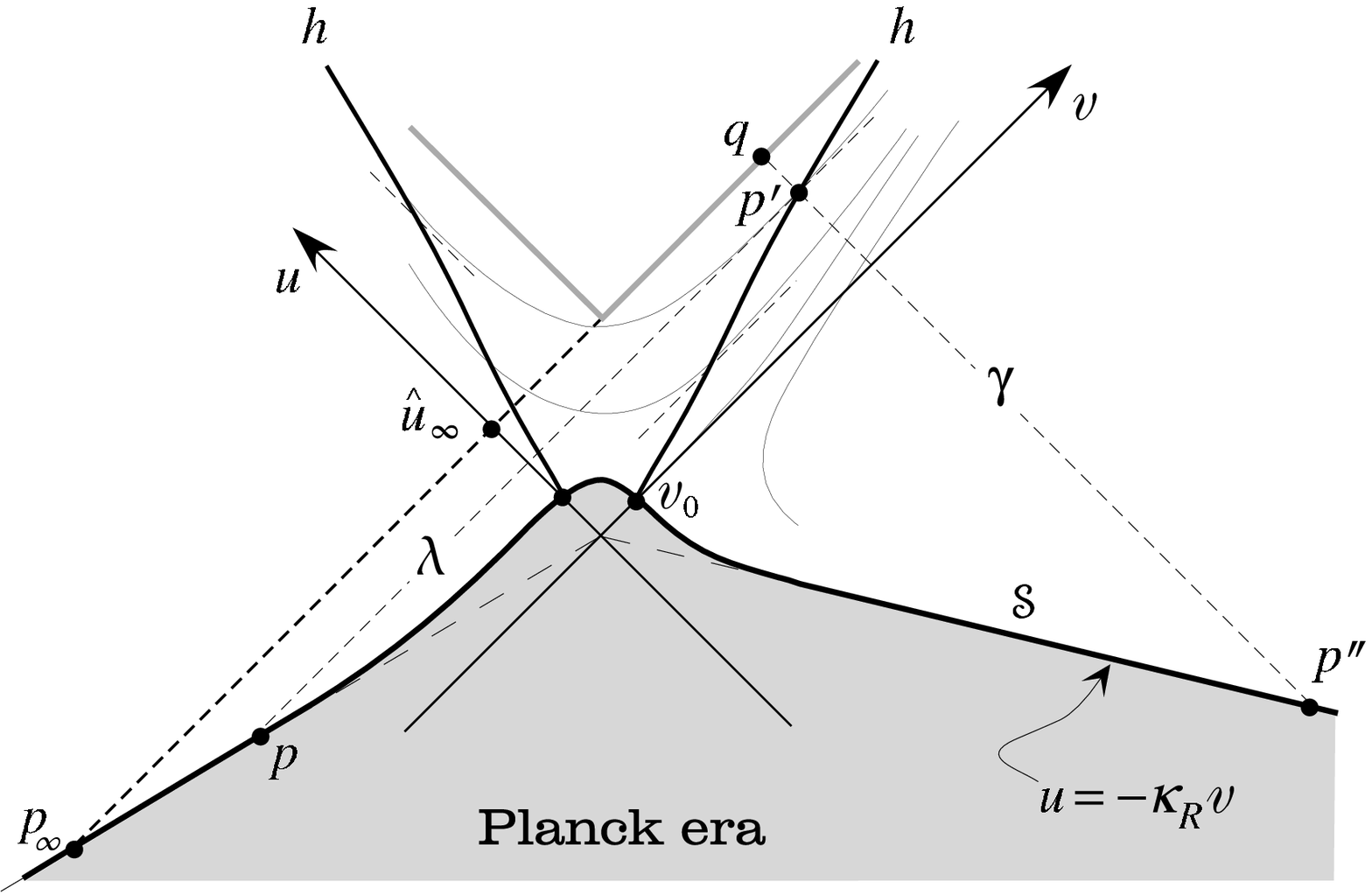}{\label{fig:kartinka} The section
$\phi=\theta=0$ of the Einstein--Rosen wormhole. The thin solid lines are
surfaces $r=const$. The gray angle is the event horizon.}
such that $F$, $r$, and their first derivatives are equal, on \s , to $
\mathring{F}$, $\mathring{r}$, and their  derivatives, respectively. The
surface is subject to the following three requirements:
        \begin{sv-va}
        \item It is spacelike between the horizons,  i.~e.\
        at\label{sv:sp} $x< 1$;
        \item For the points of \s\ with $x>1$ and $u>v$ the dependence
        $u(v)$ is a smooth positive function without maximums.
        The same must hold also with $v$ and $u$
        interchanged;\label{sv:vic}
        \item Far from the wormhole (i.~e.\
        at $r\gg m_0$) \s\ must be just a surface of  constant
        Schwarzschild time, that is it must be given by the equation
        $u/v=const$.\label{sv:far}
Thus (see figure~\ref{fig:kartinka}) for any point $p\in\s$ with $r(p)\gg
m_0$
\begin{equation*}
\begin{aligned}
v(p)>u(p) &\quad\hence\quad u(p)=-\kappa_R v(p),
\\ v(p)<u(p) &\quad\hence\quad  v(q)=-\kappa_L u(q),
\end{aligned}
\qquad\kappa_R,\kappa_L>0.
\end{equation*}
        \end{sv-va}
Condition \eqref{sv:sp} restricts substantially the class of wormholes
under examination, in contrast to \eqref{sv:vic}, which is of minor
importance and can be easily weakened, if desired.  The idea behind
\eqref{sv:far}
 is that far from the wormhole the Schwarzschild time
becomes the `usual' time and that the Planck era ended
--- by that usual time --- simultaneously in different regions of the
universe. Though, remarkably, \eqref{sv:far} does not affect the relevant
geometrical properties of \os, it proves to be very useful in their
interpretation. In particular, it enables us to assign in an intuitive
way the `time' $T$ to any event $p'$ near the throat of the wormhole.
Namely, $p'$ happens at the moment when it is reached by the photon
emitted in the end of Planck era from the point $p$ (or $p''$) located in
the left (respectively, right) asymptotically flat region. The distance
from this point to the wormhole
--- when it is large enough --- is approximately $ 2m_0\ln
u^2(p')\kappa_L$ (respectively, $\approx 2m_0\ln v^2(p')\kappa_R$).
Taking this distance to be the measure of the elapsed time from the end
of the Planck era we define
\begin{equation}\label{eq:dist}
T_L(p')= 2m_0\ln u^2(p')\kappa_L,\qquad
 T_R(p')=2m_0\ln v^2(p')\kappa_R
\end{equation}
(even though $\nabla T$ is null, as is with `advanced' and `retarded'
time in the Schwarzschild case). Note that as long as we consider the two
asymptotically flat regions as different and totally independent (i.~e.\
up to section~\ref{sec:conc}) there is no relation between $\kappa_R$ and
$\kappa_L$, nor there is a preferred value for either of them.

Among other things the choice of \s\ fixes the coordinates $u$ and $v$ up
to a transformation
\begin{equation}\label{eq:Ctransf}
{u }\mapsto u'=C{u },\quad {v}\mapsto v'=C^{-1}{v }.
\end{equation}
To fix this remaining arbitrariness and thus to make formulas more
compact  we require
\label{p:calibr}
\begin{equation*}
u_0=v_0,
\end{equation*}
where $u_0$ and $ v_0$ are the coordinates of the intersections of \s\
with the coordinate axes, see figure~\ref{fig:kartinka}. Though no
reasons are seen to think that wormholes with some particular values of
$v_0$ are more common than with any other, we restrict our consideration
to those with
\begin{equation*}1<\varpi <\tfrac{\sqrt 5 +1}{2},\qquad\text{where}\quad
\varpi  \equiv
e\eta/v_0^2,\qquad\eta \equiv 16\pi Km_0^{-2}.
\end{equation*}
As we shall see below the wormholes with smaller $\varpi$ may be
non-traversable, while those with larger $\varpi$ evaporate too intensely
and cannot be studied within our simple model. To summarize, we have four
independent parameters $m_0$, $\varpi$, and $\kappa_{R(L)}$, all values
of which are considered equally possible as long as $m_0\gg 1$, $\varpi
\in (1,\tfrac{\sqrt 5 +1}{2})$, and $\kappa_{R(L)}>0$.

Our subject will be the (right, for definiteness) horizon, by which I
understand the curve $\hor$ lying in the $(u,v)$-plane and defined by the
condition
\begin{equation}\label{eq:rv0}
r,_v\ogr{}{\hor}=0.
\end{equation}
By  \eqref{eq:rvs} $r,_v$ is negative in all points of \s\ with $u>0$ and
vanishes in the point $(0,v_0)$. In this latter point the horizon starts.
$\hor$ cannot return to \s, because there are no more points $r,_v=0$ on
\s\ [by condition~\eqref{sv:vic}]. Neither can it have an end point,
being a level line of the function with non-zero [by
condition~\eqref{eq:znakop} imposed below] gradient. So, $\hor$ goes from
\s\ to infinity
 dividing the plane $(u,v)$ above \s\  into two
parts:  $r,_v$ is strictly negative to the left of $\hor$ and strictly
positive to the right. So the horizon exists and is unique. The physical
meaning of $h$ is that its each small segment shows where the event
horizon would pass if the evolution `stopped at this moment'. The metric
in that case would be just the Schwarzschild metric with  mass
\begin{equation}\label{eq:def_mwh}
\mwh(v)\equiv \tfrac{1}{2}  r(\hor(v)).
\end{equation}
The fact that $\hor$ can be parametrized by $v$, as is implied in this
expression, will become obvious below. Alternatively the horizon can be
parametrized by $\mwh$.

\paragraph{Notation.} From now on we shall write $\oh
f$ for the restriction of a function $f(u,v)$ to $\hor$. In doing so we
view $\oh f$ as a function of $v$ or $\mwh$ depending on which
parameterization is chosen for \hor\ (this is a --- slight --- abuse of
notation, because strictly speaking $\oh f(v)$ and $\oh f(\mwh)$ are
different functions, but no confusion must arise). Partial derivatives
are, of course, understood to act on $f$, not on $\oh f$. Thus, for
example,
\[
\mwh= \tfrac{1}{2} \oh r, \qquad
\frac{\partial}{\partial v}\oh r,_u = r,_{uv}( \hor(v)), \qquad
\oh \varphi,_{uv}(\mwh)=\varphi,_{uv}(\hor(\mwh)), \qquad \text{etc.}
\]
In conformity with this notation the function $v\to u$ whose graph is
$\hor$ will be denoted by $\oh u(v)$, while $\oh u(\mwh)$ is a shorthand
notation for $\oh u(v(\mwh))$.
\par\medskip\par
Traversability of the wormhole is determined by the fact that $\oh{u}
(\mwh)$ tends to $\oh{u}_\infty>v_0$ as $\mwh\to 1$ (what happens at
smaller $\mwh$ is, of course, beyond the scope of this paper). Indeed,
consider a null geodesic $\lambda$ given by $u=u(p')$, where $p'\in h$.
In our model $\oh{r},_{vv}$ is strictly positive [see
eq.~\eqref{Re_eqs:vv_H} below] and hence $\lambda$ intersects $\hor$ once
only. As we have just discussed $r,_v$ is negative in all points of
$\lambda$ preceding $p'$ and is positive afterwards. So, $r$ reaches in
$p'$ its minimum on $\lambda$. That is the photon emitted in
$p=\lambda\cap\s$ passes in $p'$ the throat of the wormhole\footnote{I
call it a throat just because it is the narrowest place on the photon's
way, but, since $\lambda$ is orthogonal to the sphere $u,v=const$ through
$p'$, this term is in agreement with what is proposed in
\cite{HocVis}.} and
\emph{escapes to infinity}. As we move from $p$ to the left the same
reasoning applies to all photons as long as their $u$-coordinates are
small enough to enforce the intersection of $\hor$ and $\lambda$. The
boundary of this region is generated by the points $p_\infty$  with
\[u(p_\infty)=\oh{u}_\infty\equiv \sup_{\mwh\in(1,m_0)}\oh{u}(\mwh)
\]
(as we shall see the supremum is provided, in fact, by $\mwh=1$).
Correspondingly, we define the \emph{closure time} ---  the moment when
the wormhole ceases to be traversable for a traveler wishing to get from
the left  asymptotically flat region to the other one:
\begin{equation*}
T_\text{L}^\text{cl}\equiv 2m_0\ln \oh{u}_\infty^2\kappa_{L}.
\end{equation*}
Similarly is defined the
\emph{opening time} $T_\text{L}^\text{op}\equiv 2m_0\ln \oh{u}_0^2\kappa_{L}$.
So, the time of traversability of a wormhole is
\begin{equation}\label{Ttrav}
T_\text{L}^\text{trav}=T_\text{L}^\text{cl}-T_\text{L}^\text{op}=
 4m_0\ln\frac{\oh{u}_\infty}{\oh{u}_0}.
\end{equation}
Thus the goal of the paper is essentially to estimate
$\oh{u}_\infty/{\oh{u}_0}$.

\rem The fact that $r\geq r(p')$ for all points of $\lambda$, guarantees that within our model
no photon from the singularity $r=0$ will come out of the wormhole. So,
in spite of evaporation and the WEC violations involved, the wormhole
fits in with  the (weak) cosmic censorship conjecture.

\subsection{Weak evaporation assumption.}\label{sec:WE}
The physical assumption lying in the heart of the whole analysis is the
``evaporation stability" of the Einstein--Rosen wormhole, i.~e.\ I assume
that there is a solution of the system (Einstein equations + field
equations) which starts from \s\ and has the following property: the
geometry in a small neighbourhood of any point $p$ is similar to that in
a point $\mathring p$ of the Schwarzschild space with mass $\mathring m$
(of course $\mathring p$ and $\mathring m$ depend on $p$), while the
stress-energy tensor in $p$ is small and close to $\Un T
_{\mu\nu}(\mathring m, x(\mathring p))$.

More specifically I require of \os\ that to the future of \s
\begin{subequations}
\begin{equation}\label{eq:znakop} r,_{uv}<0
\end{equation}
[cf.~\eqref{eq:xuv}] and
\begin{equation}\label{eq:neor}
  \nabla r\neq 0.
\end{equation}
\end{subequations}
The latter means that at \s\ the throat of the wormhole is already
contracting and that later contraction does not pass into expansion.

The requirement that  $T _{\mu\nu}$ in a point $p\in \hor$ is close to
$\Un T _{\mu\nu}(\mwh, 1)$  is embodied in the assumption that  the
relations
\eqref{eq:TonSchor_S} are valid when the sign $\mathring{}$ is removed and
$m_0$ is replaced with $\mwh$:
\begin{align}\label{eq:tvv_hor}
 \oh{r},_{u }^{2}\oh{  T}_{v v } &  =
-\tfrac{K}{16}\oh{F}^4\mwh^{-4},\qquad 0<K \ll 1;
  \\\label{eq:Fsuu}
4\pi \oh{r},_{u }^{-2} \oh{T}_{u u }&=c\mwh^{-4},\qquad c \ll 1;
 \\\label{eq:prenuv}
4\pi |\oh{T}_{vu}|&\ll\tfrac{1}{16}\oh{F}^2\mwh^{-2}.
\end{align}
I also assume that outside the horizon
\begin{equation}\label{eq:uupos}
  T_{uu}\geq 0.
\end{equation}
In the Schwarzschild case this inequality is known to hold at $x\approx
1$, see~\eqref{T0uu}. Elster's results ($\Un{T}_{uu}\sim \mu +p_r + 2s$
in notation of
\cite{Elster}) make it obvious that
\eqref{eq:uupos} holds also at $x>1.5$. It is still possible, of course,
that $\Un{T}_{uu}$ by whatever reasons changes its sign
somewhere\footnote{This hopefully can be verified numerically. Some
arguments against this possibility can be found in
\cite{viss:Unr}.} between 1 and 1.5, however, even if \eqref{eq:uupos}
breaks down the results established below are not affected unless the
violation is so strong that it changes the sign of the relevant
\emph{integral}, see \eqref{eq:ru/F2}. Finally, I assume
that
\begin{equation}\label{eq:prenthth} |T_{\theta\theta}|\ll
\tfrac{1}{2\pi }r|r,_{vu}|F^{-2}.
\end{equation}
Again, the corresponding inequality in the Schwarzschild case --- it is
$2\tau_4\ll  m_0 ^2/x$, see
\eqref{eq:strSch} and \eqref{eq:xuv} --- holds both on the horizon and at large
$x$, see eqs.~\eqref{eq:tau4_hor} and \eqref{eq:tau4_inf}. And, again, we
actually do not need \eqref{eq:prenthth} to be true  \emph{pointwise}.
The smallness of the relevant integral [see eq.~\eqref{eq:phi,u}] would
suffice.

\rem All these assumptions are local in the sense that to check their
validity  an observer in a point $p$ does not need to know anything
beyond a small vicinity of $p$. For, the requirement that the metric in
this vicinity is (approximately) \eqref{eq:Sch} fixes the coordinates up
to the transformation \eqref{eq:Ctransf} and the assumptions are
invariant under that transformation.

\subsection{Groundless apprehensions}

Now that the model is built finding out whether the Einstein--Rosen
wormhole is traversable becomes a  matter of mathematics. But
traversability of wormholes, let alone the evolution of the black hole
horizons, are long being investigated and both theories have arguments
that seem to enable one to answer in the negative even without solving
any equations. In this subsection I point the holes in two of these
conceivable arguments.

\subsubsection{Quantum inequalities}

From \eqref{eq:tvv_hor} it is seen that the weak energy condition is
violated in some \emph{macroscopic} region $\mathcal V$ around the throat
of the wormhole. At the same time the energy density $\rho$ measured by a free falling observer
--- whose proper time we denote by $t$ --- obeys in
$\mathcal V$ the inequality
\cite{singed}
\begin{equation}\label{QI}
\int_{t_1}^{t_2}\rho(t)\,\rmd t\lesssim |t_1-t_2|^{-3},
\qquad\text{when}\quad |t_1-t_2|\lesssim m_0.
\end{equation}
The combination of these two properties in a few occasions (note that the
global structure of the spacetime is irrelevant here, it need not be a
wormhole) led to quite impressive estimates. Thus, in particular, it was
found in
\cite{PFWD} that in the Alcubierre bubble and in the Krasnikov tube there
are three-surfaces $\Xi$ and unit timelike vector fields $\boldsymbol u$
such that
\begin{equation}\label{Etot}
\int_{\Xi} T_{\mu\nu}u^\mu u^\nu\,\rmd^3 V
\approx - 10^{32} M_\text{Galaxy}.
\end{equation}
The figure in the right hand side is so huge that both spacetimes were
dismissed as `unphysical'. So, is there not any danger of that kind in
our case?

The answer is negative by at least two reasons. First, we explore not
 the capabilities of a hypothetical advanced
civilization (as is usual in discussing the above-mentioned spacetimes),
but a natural phenomenon. And there is a vital interpretational
difference between these two situations.  Indeed, in the former case the
fact that a physical quantity has a presumably unrealistic value can be
used as a ground for ruling the corresponding solution out as unphysical
or unfeasible. But in the case at hand the situation --- once the
assumptions about the initial data, the values of the parameters, and the
other constituents of the model are admitted reasonable --- is
\emph{reverse}. If calculations yield \eqref{Etot}, this would not signify
that the spacetime is unphysical. On the contrary, it would mean that
huge values of the integral may occur in physically appropriate
situations and thus cannot serve as sign of unfeasibility of a spacetime.
%
%

Second, the estimates like
\eqref{Etot} do not follow from
\eqref{QI} automatically. Additional assumptions are necessary and the
approximate equality
\begin{equation}\label{R2G}
\frac{\max |G_{\hat k\hat l}(p)|}{
  \max |R_{\hat \imath\hat \jmath\hat m\hat n}(p)|}
 =\frac{8\pi\max |T_{\hat k\hat l}(p)|}{
  \max |R_{\hat \imath\hat \jmath\hat m\hat n}(p)|}
\approx 1
  \qquad p\in \mathcal V.
\end{equation}
is among  them \cite{NoQI}. At first glance, violation of this equality
would signify some unnatural fine-tuning (note that $\approx$ can be
understood quite liberally, the difference in 5--10 orders being
immaterial). In fact, however, this is not the case:  eq.~\eqref{R2G}
corresponds to the situation in which the
  geometry of
$\mathcal V$ is defined mostly by its (exotic) matter content, while the
contribution to the curvature of the Weyl tensor is neglected. But in
four dimensions this is not always possible. For example, eq.~\eqref{R2G}
breaks down, in
\emph{any} non-flat empty region (the numerator vanishes there
but the denominator does not). And the Einstein--Rosen wormhole is just
another example. Loosely speaking, the Schwarzschild spacetime is a
wormhole by itself. In making it traversable exotic matter is needed not
to shape the spacetime into a wormhole, but only to keep the latter ajar.

\subsubsection{The gap between the horizons}

The model built above is not entirely new. The  behaviour of the apparent
horizon in very similar assumptions was studied back in 1980s (see,
i.~e.\ \cite{obz} for some review). The spacetime under consideration,
though, was not the wormhole
\os, but the black hole originating from gravitational collapse (such a
spacetime is not a wormhole, nor is it empty). The general consensus (see
though \cite{Hajicek}) was that the backreaction results only in the
shift of the event horizon to a radius smaller than $2m$ by $\delta\sim
m^{-2}$, which is physically negligible~\cite{Bardeen}. To see why such
an overwhelmingly small  $\delta$ does not make wormholes non-traversable
note that $\delta$ is the shift in radius and
\emph{not the distance} between the horizons\footnote{The event horizon
is a null surface and there is no such thing as the (invariant) distance
between a point and a null surface. Consider, for example, the surface
$t=x +\delta$ in the Minkowski plane.\label{f-n} Is it far from, or close
to the origin of coordinates? The answer is: neither. Simply by an
appropriate coordinate transformation one can give \emph{any} value to
$\delta$.}. That is $\delta=r(q)-r(p')$, see figure~\ref{fig:kartinka}.
Clearly this quantity has nothing to do with traversability of the
wormhole.

\section{The evolution of the horizon}
The Einstein equations for the metric \eqref{eq:ourmetr} read
\begin{align}   4\pi T_{vu} & =\frac{F^2}{4r^2} + \label{eqs:vu}
   (rr,_{vu}+ r,_vr,_u)r^{-2}\\
   \label{eqs:deistvv}
  4\pi T_{vv} & = (2r,_v\varphi,_v-r,_{vv})r^{-1}
  \\
  \label{eqs:vv}
  4\pi T_{uu} & = (2r,_u\varphi,_u-r,_{uu})r^{-1}
  \\ &\hspace{6em}\qquad = -\frac{F^2}{r}\Big(\frac{r,_u}{F^2}\Big),_u
   \label{eqs:vv'}\tag{\ref{eqs:vv}$'$}
   \\
  4\pi T_{\theta\theta} & = -\frac{2r^2}{F^2}(r,_{vu}/r + \varphi,_{vu}).
  \label{eqs:thth}
\end{align}
On the horizon  the left hand side in
\eqref{eqs:vu} can be neglected by \eqref{eq:prenuv}, while $r,_v$
vanishes there by
definition and we have
\begin{equation}\label{Re_eqs:vu_H}
\oh r,_{vu}=-\frac{\oh{F}^2}{8\mwh}.
\end{equation}
Eqs.~\eqref{eqs:deistvv} and \eqref{eq:tvv_hor} give
\begin{equation}\label{Re_eqs:vv_H}
\oh r,_{vv}=\frac{\pi K \oh{F}^4}{2\mwh^3\oh{r},_u^2 }.
\end{equation}
Likewise,  \eqref{eqs:vv} and
\eqref{eq:Fsuu} result in
\begin{equation}\label{Re_eqs:vv}
\oh r,_{uu}=2\oh{r},_u\oh \varphi,_u- 2c \oh{r},_u^2\mwh^{-3}.
\end{equation}
Finally, equations \eqref{eqs:thth} and \eqref{eq:prenthth} yield
\begin{equation}\label{Re_eqs:thth}
\varphi,_{vu}=-r,_{vu}/r.
\end{equation}

\subsection{$\boldsymbol{\oh{u}}$ as function of $\boldsymbol \mwh$}
Our aim in this subsection is to find the function $\oh{u}(\mwh)$. To
this end we, first, use eqs.~\eqref{Re_eqs:vu_H}--\eqref{Re_eqs:thth} to
find a system of two ODE defining $\oh{u}(\mwh)$ [these are
eqs.~\eqref{eq:dmdu} and
\eqref{eq:ur4ru}, below]. Then for wormholes with
\begin{subequations}\label{eq:ogrnapi}
\begin{align}\label{eq:ogr1}
\varpi &>1
\\
\label{eq:ogr1'}
\varpi &<\tfrac{\sqrt 5 +1}{2}
\end{align}
\end{subequations}
we integrate the system and obtain a simple explicit expression for
$\oh{u}$.

The horizon can be parametrized by $v$, or by $\mwh$ (as was already
mentioned), or finally by $u$. The relations between the three
parameterizations are given by the obvious formulas:
\begin{equation}\label{eq:dmdu}
2\frac{\rmd \mwh}{\rmd u}=\frac{\rmd \oh{r}}{\rmd u} =\oh{r},_u
\end{equation}
and
\begin{equation}\label{eq:Y}
\frac{\rmd v}{\rmd u}=-\frac{\oh r,_{vu}}{\oh r,_{vv}},
\end{equation}
of which the former follows right from the definitions \eqref{eq:rv0},
\eqref{eq:def_mwh} and the latter from the fact that $0=
\rmd\oh{r},_v=\oh r,_{vu}\rmd u +\oh r,_{vv}\rmd v$ on  $\hor$.
These formulas enable us to write down
\begin{equation} \label{eq:whi}
\frac{\rmd}{\rmd \mwh} \oh{r},_u=
\frac{\rmd u}{\rmd \mwh}
 \left(\frac{\partial}{\partial u}
+\frac{\rmd v}{\rmd u} \frac{\partial}{\partial v}\right)\oh r,_u=
2\oh{r},_u^{-1}\Big(\oh r,_{uu}-\frac{\oh r,_{vu}^2}{\oh r,_{vv}}\Big).
  \end{equation}
Using \eqref{Re_eqs:vv} and the relation
\begin{equation*}
\frac{\oh r,_{vu}^2}{\oh r,_{u}^2\oh r,_{vv}}=
\frac{\mwh }{32\pi K},
\end{equation*}
which follows from eqs.~\eqref{Re_eqs:vu_H} and \eqref{Re_eqs:vv_H}, one
can rewrite \eqref{eq:whi} as
\begin{equation}\label{eq:ln_prom}
\oh{r},_u^{-1}  \frac{\rmd \oh{r},_u}{\rmd \mwh}=
4\frac{\oh \varphi,_u}{\oh{r},_u}-4c\mwh^{-3}
 - \frac{\mwh }{16\pi K}.
\end{equation}
To assess the first term in the right hand side consider the segment
$\lambda$ of the null geodesic $u=const$ between a pair of points $p\in
\EuScript S $, $p'\in \hor $. Below I write for brevity  $\bar{r}$,
$\bar{x},_u$, etc.\ for  $r(p)$, $x_u(p)$, etc. (note that in this
notation $\bar{u}=\oh{u}$). By
\eqref{Re_eqs:thth} and
\eqref{eq:phiu}
\begin{equation}\label{eq:phi,u}
\oh\varphi,_u=
 \varphi,_u(p')=\varphi,_u(p)+\int_\lambda\varphi,_{uv}\,\rmd v
=
-\frac{1+\bar{x}}{2\bar{x}}\bar{x},_u-
\int_\lambda \frac{r,_{uv}}{r}\,\rmd v.
\end{equation}
The sign of $r,_{uv}$ is constant by \eqref{eq:znakop}, while $r$ --- as
was shown in section~\ref{sec:monot} it monotonically falls on $\lambda$
--- varies from $\bar{r}$ to $2\mwh $. Thus,
\begin{gather}
\oh \varphi,_u=\big(\tfrac{1}{2m_*} - \tfrac{1+1/\bar{x}}{4m_0}\big)\bar{r},_u
-\tfrac{1}{2m_*}
\Big(\bar{r},_{u}+ \int_\lambda r,_{uv}\,\rmd v \Big)
=\big(\tfrac{1}{2m_*} - \tfrac{1+1/\bar{x}}{4m_0}\big)\bar{r},_u
-\tfrac{1}{2m_*}
\oh r,_{u}\nonumber
\\
\label{eq:def_m*}
\mwh  \leq m_*\leq \tfrac{1}{2}\bar{r}.
\end{gather}
Substituting this in \eqref{eq:ln_prom}  and neglecting the terms $\sim
m_*^{-1}$, $\mwh^{-3}$ in comparison with the last term we finally get
\begin{equation}\label{eq:ur4ru}
\oh{r},_u^{-1}\frac{\rmd \oh{r},_u}{\rmd \mwh} =\frac{2\xi \bar{x},_u}{\oh{r},_u}-
\frac{\mwh}{16\pi K},
\qquad
\xi\equiv (\tfrac{2m_0}{m_*} - 1-
\tfrac{1}{\bar{x}}),
\quad
\oh{r},_u(m_0)=-2\frac{m_0v_0}{e}
\end{equation}
(the last equation follows from \eqref{eq:rus} and serves as a boundary
condition for the differential equation). Introducing new notations
\[
\mu\equiv\frac{\mwh }{m_0},  \qquad
y(\mu)\equiv e^\frac{1-\mu^2}{2\eta }
\]
one readily finds the solution of this equation:
\begin{equation}\label{eq:resh}
\oh{r},_u(\mu)= -2\frac{m_0v_0}{e}[1+\Xi(\mu)]\, y(\mu),
\qquad
\Xi\equiv
\frac{e}{v_0}
\int^{1}_{\mu }\frac{\xi }{ y}
\left(\frac{\bar{x}-1}{\bar{u}\bar{x}}\right)\,\rmd \mu
\end{equation}
and, correspondingly, [the first equality is an obvious consequence of
eq.~\eqref{eq:dmdu}]
\begin{equation}\label{eq:u(m)}
  \oh{u}(\mwh)=2\int_{m_0}^{\mwh}\frac{\rmd
  \mwh}{\oh{r},_u}
=A(\mu)\frac{e}{v_0}\int_\mu^1\frac{\rmd\zeta}{y(\zeta)}.
\end{equation}
Here $A(\mu)$ is an unknown function  bounded by
\begin{equation}\label{eq:boundsA}
[\max_{(\mu,1)} (1+\Xi)]^{-1}\leq  A(\mu)\leq [ \min_{(\mu,1)}
(1+\Xi)]^{-1}.
\end{equation}

In the remainder of this subsection I demonstrate that $|\Xi|<1$, which
implies, in particular, that $\oh{u}(m)$ monotonically falls and
therefore $\oh{u}_\infty$ is just $\oh{u}(1)$. To simplify the matter the
further consideration will be held separately for small and for large
$\oh u$.
\paragraph{The case
$\boldsymbol{\oh{u} < v_0}$.} On this part of  $\hor$ it is possible that
$\lambda\cap\s$ consists of one, two, or three points. But  one of them
always lies between the horizons and it is this point that we take to be
the point $p$ that enters \eqref{eq:def_m*} and thus
\eqref{eq:u(m)}. We then are ensured that $\bar{x}<1$ and $\bar{v}<v_0$.
By
\eqref{eq:strSch'} it follows
\begin{gather*}
(1-\bar{x})/\bar{u}=\bar{v}e^{-\bar{x}}<v_0
\\
\bar{x}> 1 - (1-\bar{x})e^{\bar{x}} = 1 - \bar{u}\bar{v} > 1-v_0^2
\end{gather*}
and therefore (recall that $\eta\ll 1$ and by
\eqref{eq:ogr1} so is $v_0$)
\begin{equation}\label{eq:xi@U<v0}
\frac{|\bar{x}-1|}{\bar{u}\bar{x}}<2v_0.
\end{equation}
Now note that by \eqref{eq:def_m*} at $\bar x <1$
\[1\leq\frac{1}{\bar{x}}\leq\frac{m_0}{m_*}\leq\frac{1}{\mu}\]
and hence
\begin{equation}\label{eq:Ocnaxi}
0<\xi=(\tfrac{m_0}{m_*} - 1)+ (\tfrac{m_0}{m_*}-
\tfrac{1}{\bar{x}})\leq
2\frac{1-\mu}{\mu}.
\end{equation}
Consequently,
\begin{equation}
|\Xi|\leq
\frac{2e}{v_0}
\int^{1}_{\mu}\frac{1-\zeta }{\zeta y(\zeta)}
\frac{|\bar{x}-1|}{\bar{u}\bar{x}}\,\rmd \zeta
 <4e\int^{1}_{\mu}\label{eq:Xi_ass}
\frac{1-\zeta}{\zeta y(\zeta)}\,\rmd \zeta.
\end{equation}
To proceed let us write down the following equality obtained by
integrating by  parts
\[
\int^{1}_{ \mu }
\frac{1-\zeta}{\zeta y(\zeta)}\,\rmd \zeta
=\eta \left[-N+ e^{-\frac{1}{2\eta}}
\int^{1}_{ \mu }
\Big(\frac{2}{\zeta^3 }-\frac{1}{\zeta^2}\Big)
e^\frac{\zeta^2}{2\eta}\,\rmd\zeta\right]
 ,\qquad N\equiv\frac{1- \mu}{{\mu }^2 }e^\frac{-1+{ \mu }^2}{2\eta} .
\]
Note that the integrand in the right hand side monotonically grows at
$1/m_0\leq \zeta<1$ (i.~e.\ as long as the wormhole remains macroscopic).
So, splitting when necessary (i.~e.\ when $\mu< 1-100\eta$) the range of
integration by the point $\zeta=1-100\eta$ and replacing the integrand on
either interval by its maximum we obtain the following estimate (recall
that $100\eta\ll 1$)
\[
e^{-\frac{1}{2\eta}}\int^{1}_{ \mu }
\Big(\frac{2}{\zeta^3 }-\frac{1}{\zeta^2}\Big)
e^\frac{\zeta^2}{2\eta}\,\rmd\zeta
\leq
\Big(\frac{2}{\mu^3
}-\frac{1}{\mu^2}\Big)e^\frac{\mu^2-1}{2\eta}\ogr{}{\mu=1-100\eta}+
100\eta
\approx  e^{-100} + 100\eta.
\]
So, taking into consideration that $N$ is positive,
\begin{equation}\label{eq:ocInt}
Z\equiv\int^{1}_{ \mu }
\frac{1-\zeta}{\zeta y}\,\rmd \zeta \leq  e^{-100}\eta + 100\eta^2
\qquad\forall\,\mwh >1.
\end{equation}
Substituting which in \eqref{eq:Xi_ass} yields $|\Xi|\ll 1$ and hence
$A(\mu)=1$. Correspondingly,
\begin{equation*}\label{eq:HorMalU}
\oh{u}=\frac{e}{v_0}\int_\mu^1\frac{\rmd\zeta}{y(\zeta)}.
\end{equation*}
This expression is valid on the whole segment $\oh{u} < v_0$, i.~e.\ at
$\mu\geq\mu_*$
\begin{equation*}
\mu_*\colon\quad \oh{u}(\mu_*)=v_0.
\end{equation*}
To find $\mu_*$ we employ the formula (see, e.~g.,
\cite{Fed})
\[\int^\frac{\mu}{\sqrt{2\eta}}_0 e^{\zeta^2}\rmd \zeta=
\frac{\sqrt{\eta/2}}{\mu}\:e^{\frac{\mu^2}{2\eta}},
\]
which is valid (asymptotically) at small $\eta$.
\[
v_0=\oh{u}(\mu_*)=
\frac{e^{1-\frac{1}{2\eta}}}{v_0}
    \int_{\mu_*}^1e^{\frac{\zeta^2}{2\eta}}\rmd\zeta =
\frac{e^{1-\frac{1}{2\eta}}\eta}{v_0}(e^{\frac{1}{2\eta}}
 -\tfrac{1}{\mu_*}e^{\frac{\mu_*^2}{2\eta}})
=\frac{e\eta}{v_0}(1-\tfrac{1}{\mu_*}e^{\frac{\mu_*^2-1}{2\eta}}).
\]
So,
\begin{equation}\label{eq:mu*}
\tfrac{1}{\mu_*}e^{\frac{\mu_*^2-1}{2\eta}}=1-\varpi^{-1}.
\end{equation}

\paragraph{The case $\boldsymbol{\oh{u}> v_0}$.}
Now $\bar{x}>1$ and instead of \eqref{eq:xi@U<v0} we have
\begin{equation*}
\frac{\bar{x}-1}{\bar{u}\bar{x}}<
  \frac{1}{v_0}=\frac{\varpi}{e\eta}v_0
\end{equation*}
and instead of \eqref{eq:Ocnaxi}
\[ |\xi|\leq \frac{1-\mu}{\mu} + \frac{1}{\mu}.
\]
Substituting these inequalities in \eqref{eq:resh} and neglecting the
contribution of the segment $(\mu_*,1)$ in $\Xi$ gives
\[
|\Xi|\leq
\frac{\varpi}{\eta}\int^{\mu_*}_{\mu } y^{-1}
\left(\frac{1-\zeta}{\zeta} + \frac{1}{\zeta}\right)\,\rmd \zeta
\leq\frac{\varpi}{\eta} Z +
\frac{\varpi}{\eta}\int^{\mu_*}_{\mu } \frac{\rmd \zeta}{\zeta y}.
\]
The first term can be neglected by \eqref{eq:ocInt}  and we have
\[
|\Xi|\leq\frac{\varpi}{\eta}\int^{\mu_*}_{1/m_0 }
 e^\frac{\zeta^2-1}{2\eta }\zeta^{-1}\rmd \zeta=
\frac{\varpi}{2\eta}e^{-\frac{1}{2\eta}}
\int^{\frac{\mu_*^2}{2\eta}}_{\frac{1}{2\eta m_0^2 }}
e^\zeta\zeta^{-1}\rmd \zeta=
\frac{\varpi}{\mu_*^2}e^{\frac{\mu_*^2-1}{2\eta}}=\varpi-1
\]
(the last equality follows from \eqref{eq:mu*} and the last but one ---
from the fact that $\int_a^be^\zeta\zeta^{-1}\rmd \zeta\sim e^b/b$ at
large $b$). Thus on this segment of the horizon
\begin{gather*}
\oh{u}=v_0+A(\mu)\frac{e}{v_0}\int_\mu^{\mu_*}\frac{\rmd\zeta}{y(\zeta)}
\\
\frac{1}\varpi\leq A\leq \frac{1}{2-\varpi}.
\end{gather*}
Whence, in particular,
\begin{subequations}\label{eq:u_inf}
\begin{equation}\label{u_infMIN}
\oh{u}_\infty \geq
 v_0 +\frac{e}{v_0\varpi}\eta(1-\varpi^{-1})= v_0(2-\varpi^{-1})
\end{equation}
\begin{equation}\label{u_infMAX}
\oh{u}_\infty\leq v_0
+\frac{1}{2-\varpi}\frac{e}{v_0}\eta(1-\varpi^{-1}) =
 v_0 \left(1+\frac{\varpi-1}{2-\varpi}\right)=
\frac{v_0}{2-\varpi}.
\end{equation}
\end{subequations}

We see that $\oh{u}_\infty>v_0$ and thus the wormhole in study proves to
be traversable. Depending on the value of $\varpi$ its time of
traversability [see \eqref{Ttrav}, \eqref{eq:ogrnapi}, \eqref{eq:u_inf}]
varies from
\begin{align}\label{Ttr}
 T_\text{L}^\text{trav}=0  &\qquad \text{at } \varpi= 1\\
\intertext{to}
T_\text{L}^\text{trav}=\alpha m_0,\quad 1.3\leq\alpha\leq 3.8
   & \qquad \text{at } \varpi= \tfrac{\sqrt 5 +1}{2}\label{eq:Tinterv}
\end{align}
It should be emphasized that the upper bound on $T_\text{L}^\text{trav}$
restricts \emph{not} the traversability time of empty wormholes (nothing
in our analysis suggests that this time is restricted at all), but
 the traversability time of \emph{the wormholes obeying
\eqref{eq:ogrnapi}};
%
%
%
 it says  not that the time the wormhole is open is less
than $4 m_0$, but only that to exceed that time a wormhole would have to
have so large $\varpi$ that our model cannot describe it. To see why it
happens and why the condition
\eqref{eq:ogrnapi} has to be imposed we need to examine the form of the
horizon in more detail.

\subsection{$\boldsymbol{\oh{u}}$ as function of $\boldsymbol v$}
To  relate $\mwh$ with $v$ let us, first, combine eqs.~\eqref{eq:dmdu}
and
\eqref{eq:Y} and substitute eqs.~\eqref{Re_eqs:vu_H} and
\eqref{Re_eqs:vv_H} into the result:
\[
\frac{\rmd v}{\rmd \mwh }=
-2\oh{r},_u^{-1}\frac{ \oh{r},_{vu}}{\oh{r},_{vv}} =\frac{\oh{r},_u
\mwh^2}{2\pi K}\oh{F}^{-2},
\]
or, equivalently,
\begin{equation}\label{eq:dv/dmu}
\frac{\rmd v}{\rmd \mu^3 }=\frac{8m_0}{3\eta} \frac{\oh{r},_u}{\oh{F}^2}.
\end{equation}
Now let  $\gamma $ be a segment of a null geodesic $v=v(p')$ from $p''\in
\EuScript S$ to $p'\in \hor $.
By \eqref{eqs:vv'} on $\gamma$
\[
\Big(\frac{r,_u}{F^2}\Big),_u \rmd u =
  - \frac{4\pi r}{F^2}T_{uu}\rmd u=
  - \frac{4\pi r}{r,^2_u}T_{uu} \Big(\frac{r,_u}{F^2}\Big)\,\rmd r
\]
and hence
\begin{multline}
\frac{\oh{r},_u}{\oh F^2}(p')=\frac{r,_u}{F^2}(p'') \cdot
\exp\left\{\int_\gamma \Big( \ln\frac{r,_u}{F^2}\Big),_u \rmd u\right\}
\\ \label{eq:ru/F2}
=-\frac{v}{8m_0}
\exp\left\{-\int_\gamma
\frac{4\pi rT_{uu}\,\rmd r}{r,^2_u}\right\}
\end{multline}
(the factor at the exponent is reduced  with the use of the first
equalities in eqs.~\eqref{eq:strSch} and \eqref{eq:rus}, which are valid
in $p''$). $\gamma$ does not intersect the left horizon and therefore
$r,_u$ is negative in each of its points. So, the integration in
\eqref{eq:ru/F2} is performed in the sense of decreasing $r$. By
\eqref{eq:uupos} it follows then
\begin{equation}\label{eq:Fru}
\frac{\oh{r},_u}{\oh F^2}(v)\leq-\frac{v}{8m_0}.
\end{equation}
Substituting which in
\eqref{eq:dv/dmu} we finally obtain
\begin{equation}\label{eq:votm}
v(\mu)\geq v_0\exp\Big\{\frac{1}{3\eta}(1-\mu^3)\Big\}
\end{equation}
and, in particular,
\[
v_\infty\equiv v(\mwh=0)\geq v_0e^{\frac{1}{3\eta}}.
\]
The latter formula enables one, among other things, to bound from below
the time of evaporation [in the sense of \eqref{eq:dist}]
\begin{equation*}\label{eq:tisp}
  T_\text{R}^\text{evap}=  2m_0\ln (v^2_\infty\kappa_R)
  \geq T_\text{R}^\text{op}+ \frac{m_0^3}{12\pi K }.
\end{equation*}

Let us check now that our model is self-consistent in that the condition
\eqref{eq:neor} does hold in
\os. To this end note that it is equivalent to the condition that the
left and right horizons do not intersect, for which it is sufficient that
\begin{equation}\label{eq:h/v<1}
  \oh{u} (\mu)<v(\mu).
\end{equation}
Clearly, this condition holds for all $\oh{u}\leq v_0$, that is for all
$\mu\geq\mu_*$. At the same time $\mu<\mu_*$ implies [the first
inequality follows from
\eqref{eq:mu*}]
\begin{equation}\label{eq:1-mu}
 \mu^3-1 <
3\eta\ln (1-\varpi^{-1}) <3\eta\ln (2-\varpi ).
 \end{equation}
It is the last inequality in this chain that we need \eqref{eq:ogr1'}
for. Combining \eqref{u_infMAX}, \eqref{eq:votm}, and \eqref{eq:1-mu} we
finally see that
\[
\oh{u}/v<\oh{u}_\infty/v\leq \frac{1}{2-\varpi}
e^{\frac{\mu^3-1}{3\eta} }<1,
\]
i.~e.\  \eqref{eq:h/v<1} is satisfied and the horizons do not intersect.
\rem By the coordinate transformation $(u,v)\to (r,\tilde v)$, where
$
\tilde v\equiv 4m_0\ln v
$, one could cast the metric into
\begin{multline*}
\rmd s^2=
-F^2r,_u^{-1}\rmd v(- r,_v\rmd v+\rmd r)+ r^2(\rmd\theta^2 +
\cos^2\theta\,\rmd\phi)
\\
= \frac{F^2v}{8r,_um_0}\Big[ \tfrac{1}{2m_0}vr,_v\rmd \tilde v^2 - 2\rmd
r\rmd\tilde v\Big] + r^2(\rmd\theta^2 + \cos^2\theta\,\rmd\phi).
\end{multline*}
So, if the integral in \eqref{eq:ru/F2} is neglected and the relation
\eqref{eq:Fru} becomes therefore an equality (as in the Schwarzschild
case), the metric takes the form
\[
\rmd s^2=-(1-2m_V/r)\rmd \tilde v^2 + 2 \rmd r\rmd \tilde v +
r^2(\rmd\theta^2 + \cos^2\theta\,\rmd\phi),
\qquad m_V\equiv r\frac{2m_0-vr,_v}{4m_0}.
\]
In the vicinity of the horizon this, in fact, is the Vaidya metric,
because
\[
4m_0m_V,_u= 2m_0r,_u -v (r,_ur,_v+rr,_{uv})\ogr{}{\hor}= 2m_0\oh r,_u +
\frac{1}{4}\,v\oh F^2=0
\]
[the second equality follows from \eqref{Re_eqs:vu_H}] and hence
\[
m_V=m_V(v)=\mwh(v).
\]

\section{Traversabilty}\label{sec:conc}
A photon arriving to the wormhole (in the `left universe') after
$T_\text{L}^\text{cl}$ will never traverse it. At the same time photons
with $u<v_0$, i.~e.\ with
$T_\text{L}< T_\text{L}^\text{op}$
 (such photons exist, unless \s\ is spacelike, which is uninteresting)
cannot traverse it either: on their way to the wormhole they get into the
Planck region,  their afterlife is veiled in obscurity. And the
traversability time $T_\text{L}^\text{trav}$ turns out to be rather
small, see
\eqref{eq:Tinterv}. For the wormholes in discussion it is only $\sim
2m_0$, which is of the order of minutes even for giant black holes which
presumably can be found in the centers of galaxies. And for a stellar
mass wormhole it measures only a few microseconds. It may appear that so
small $T_\text{L}^\text{trav}$ make the  Einstein--Rosen wormholes
useless in `inter-universe communicating' even for an advanced
civilization. This, however, is not so by the reason mentioned in
footnote~\ref{f-n}. Indeed, consider a spaceship moving in the left
asymptotically flat region towards the wormhole. Suppose, at
$T_\text{L}^\text{op}$ it is at the distance $l\gg m_0$ from the mouth
and moves so fast that reaches the mouth (i.~e.\ the vicinity of the left
horizon) at $T_\text{L} \approx T_\text{L}^\text{cl}$. Then neglecting
the terms $\sim m_0/l$ and $\sim
\oh{u}_\infty/l$ it is easy to find that the travel takes $\Delta\tau
\approx 2 \sqrt{lm_0} $ by
the pilot's clock. Thus if $T_\text{L}^\text{op}$ is large enough, the
pilot may have plenty of time to send a signal through the wormhole.

Now let us consider the intra-universe wormholes. To transform our model
into one describing such a wormhole we first enclose the throat in a
surface $\tu\colon\: r=r_M\gg 2m_0$. This surface is a disjoint union of
two cylinders $\mathbb S^2\times
\text{I\!R}^1$, one of which lies in the left asymptotically flat region
and the other in the right:
\[
\tu=\tu_L\cup\tu_R,\qquad \tu_{L(R)}\colon\: r=r_M,\ v\lessgtr 0.
\]
We shall consider the spacetime outside \tu\ (which is, correspondingly,
a disjoint union of two asymptotically flat regions $M_L$ and $M_R$) as
flat. This, of course, is some inexactness, but not too grave --- in
reality the space far enough from a gravitating body
\emph{is} more or less flat. Let us fix  Cartesian coordinates in
$M_{L(R)}$ so that the $t$-axes are parallel to the generators of \tu\
and the $x_1$-axes --- to the line $t=\phi=\theta=0$. The
$x_1$-coordinates of the points of \tu\ are understood to lie within the
range $[-r_M,r_M]$ and \s\ must be the surface $t=0$. Now an
intra-universe wormhole is obtained by the standard procedure:
\karti{t,b}{0.5\textwidth}{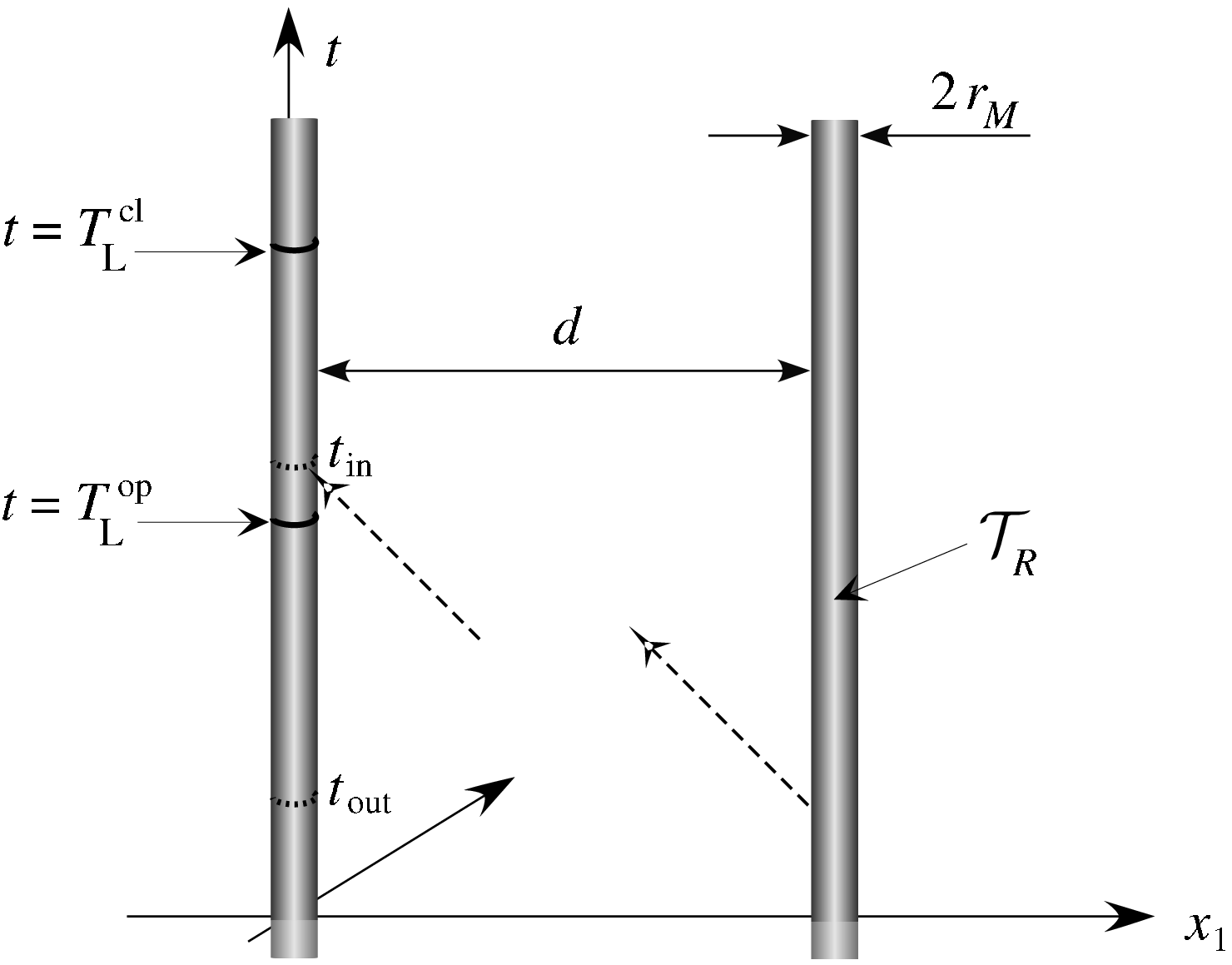}{\label{fig:intra} The two dashed lines
depict the world line of the same photon.}
one removes the regions $x_1>d/2$ from $M_L$ and $x_1<d/2$ from $M_R$ and
identify the points with the same coordinates on their boundaries (the
three-surfaces $x_1=-d/2$ and $x_1=d/2$, respectively). The resulting
spacetime, see figure~\ref{fig:intra}, is the Minkowski space in which
the interiors of two cylinders (their boundaries are $\tu_L$ and $\tu_R$)
are replaced by a connected region, so that, for example, a photon
intersecting $\tu_L$ at a moment $t_\text{in}\in
(T_\text{L}^\text{op},T_\text{L}^\text{cl})$
 emerges from $\tu_R$
at some $t_\text{out}(t_\text{in})$.

Now note that it would take only $d$ for the photon to return to $\tu_L$.
So \os\ is causal if and only if
\[
 t<t_\text{out}(t)+d\qquad \forall\,t\in
(T_\text{L}^\text{op},T_\text{L}^\text{cl}).
\]
By changing $\kappa_L$ to $\kappa'_L$  --- all other parameters being
fixed --- one shifts the interval
$(T_\text{L}^\text{op},T_\text{L}^\text{cl})$ and the graph of
$t_\text{out}(t)$ to the right by $\approx 2m_0\ln(\kappa'_L/\kappa_L)$,
see \eqref{eq:dist}. So, if $\kappa'_L$ is taken to be sufficiently large
the inequality breaks down. Which means that irrespective of the values
of $m_0$, $d$, $\varpi$, and $\kappa_{R}$, the intra-universe
Einstein--Rosen wormholes with sufficiently large $\kappa_{L}$ are time
machines.

\section{Conclusions}

We have studied the evolution of the spherically symmetric empty
wormhole, or to put it otherwise the backreaction of the Hawking
radiation on the (approximately) Schwarzschild metric. A few simplifying
assumptions were made, which physically reduced to the idea that the
metric and the vacuum polarization around each observer remain
approximately those of the Schwarzschild black hole. It turns out that
such a wormhole is characterized by three  parameters in addition to the
initial mass and the distance between the mouths. The explicit
calculations within this model have shown that \emph{for a macroscopic
time interval
--- \emph{its duration is determined by those parameters}
--- the wormhole is traversable}.

None of the assumptions made in this paper looks too wild, so its results
can be regarded as evidence for possibility of natural `transient'
wormholes. Obviously the existence of such wormholes would be of enormous
significance, the implications ranging from a process generating highly
collimated flashes to causality violation (or at least violation of  the
strong cosmic censorship conjecture).
\section*{Acknowledgements}
This  work was partially supported by RNP grant No.~2.1.1.6826.

\appendix
\section{Appendix}\label{app:1}
In this appendix I extract the relevant estimates on $\tau_i$ from  the
results obtained in
\cite{ChrFu} and
\cite{Elster}.  At large $r$ the radial pressure $\Un{T}^\theta_\theta
=\Un{T}_\phi^\phi$ equals (see eqs.~(2.6), (4.8), (5.5), and  (6.21) of
\cite{ChrFu}) to
\[
\Un{T}^\theta_\theta \approx \tfrac{\lambda}{16} K m_0^{-4}x^{-4},
\]
where
\begin{equation}\label{eq:K}
0<\lambda\leq 27,\quad K=\frac{9}{40\cdot 8^4\pi^2}.
\end{equation}
Correspondingly,
\begin{equation}\label{eq:tau4_inf}
\tau_4=16\pi m_0^4x^2\Un{T}^\theta_\theta
\approx  \pi\lambda K x^{-2},\qquad\text{at large } x.
\end{equation}
Near the horizon $\Un{T}^\theta_\theta$ was found  (numerically) in
\cite{Elster}, where it was denoted by $p_t$:
\begin{equation} \label{eq:ptr}
 0<\Un{T}^\theta_\theta \approx -\tfrac{1}{2}\tfrac{\rmd}{\rmd
x}\Un{T}^\theta_\theta
\lesssim 2\cdot 10^{-6}m_0^{-4}
\end{equation}
[the value of the derivative will be needed in \eqref{eq:valY}], whence
\begin{equation}\label{eq:tau4_hor}
\tau_4(1)\lesssim 10^{-4}.
\end{equation}
Further, for the conformal field the trace $\Un{T}_a^a$  is defined by
the conformal anomaly and in the Schwarzschild space
\[
T\equiv \Un{T}_a^a=\tfrac{m_0^{-4}}{3840\pi^2}\, x^6\approx 3\cdot
10^{-5}m_0^{-4} x^6,\qquad
 T'\ogr{}{x=1}\approx 2\cdot 10^{-4}m_0^{-4},
\]
see, e.~g., eq.~(4.8) in \cite{ChrFu}. So, for the quantity
$Y\equiv 
T - \Un{T}^\theta_\theta -\Un{T}_\phi^\phi$ one finds
\begin{equation}\label{eq:valY} Y\ogr{}{x=1}\approx
3\cdot 10^{-5}m_0^{-4},
\qquad
Y'\ogr{}{x=1}\approx 10^{-4}m_0^{-4}.
\end{equation}
In coordinates $t$, $r^*$
\[ t=2m_0 \ln (-{v }/{u }),\qquad r^*=2m_0 \ln (-{v }{u }),
\]
which are used in \cite{ChrFu,Elster}, the  Schwarzschild metric
\eqref{eq:Sch} takes the form
\[
 \rmd s^2=\tfrac{x-1}{x} \big(-\rmd t^2 + \rmd r^*{}^2)
 +\mathring{r}^2(\rmd\theta^2 + \cos^2\theta\,\rmd\phi)
 \]
and one has
\[ \Un{T}_{uv}=\tfrac{4m_0 ^2}{v u}(\Un{T}_{r^*r^*}-\Un{T}_{tt})=
-\tfrac{4m_0 ^2e^{-x}}{x}(\Un{T}_{r^*}^{r^*}+\Un{T}_{t}^t)= -\tfrac{4m_0
^2e^{-x}}{x}Y\ogr{}{x=1}\approx -4\cdot10^{-5}m_0^{-2}.
\]
From whence it follows
\begin{equation}\label{eq:tau3}
|\tau_3(1)|\approx 5\cdot 10^{-4}.
\end{equation}
Likewise,
\[
\Un{T}_{{v }{v }}=\tfrac{4m_0 ^2}{{v }^2}(\Un{T}_{tt} + \Un{T}_{r^*r^*} +2\Un{T}_{tr^*})
=\tfrac{(x-1)}{x}
\left(\tfrac{4m_0 ^2}{xe^x}\right)^2{\mathring r},_{u }^{-2}
(-\Un{T}_t^t + \Un{T}_{r^*}^{r^*} +2\Un{T}_t^{r^*}).
\]
At the horizon $x=1$ and $G(1)=H(1)=Q=0$ (see~\cite[section 2]{ChrFu} for
the definitions of the relevant functions). So  the only contribution to
$\Un{T}^{a}_b$ comes from its divergent part $T^{(2)a}_b$:
\[ T^{(2)a}_b=\frac{K}{4m_0^4x(x-1)}E^a_b,
\]
where I defined
\[
E^{t}_t=-E^{r^*}_t=E_{r^*}^t=-E^{r^*}_{r^*}=1.
\]
 Thus
\begin{equation}\label{eq:Tvv}
\Un{T}_{{v }{v }}\ogr{}{x=1}
=-16e^{-2}K{\mathring r},_{u }^{-2}.
\end{equation}
Finally,
\begin{multline}\label{T0uu}
\Un{T}_{uu}=\tfrac{x}{x-1}{\mathring r},_{u }^{2} (-\Un{T}_t^t + \Un{T}_{r^*}^{r^*}
-2\Un{T}_t^{r^*})=
\tfrac{x}{x-1}{\mathring r},_{u }^{2}\Big(
\tfrac{2x}{r^2(x-1)}(H+G)-Y
\Big)
\\
=\tfrac{x}{x-1}{\mathring r},_{u }^{2}\Big(
\tfrac{2x}{r^2(x-1)}\cdot\tfrac{1}{2}\int_{2m_0}^r
[(r'-m_0)T + (r'-3m_0)(T-2Y)]
\,\rmd r'-Y \Big)
\\
=\tfrac{x}{x-1}{\mathring r},_{u }^{2}\Big(
\tfrac{1}{x(x-1)}\int_{1}^x
[(x'-\tfrac{1}{2})T + (x'-\tfrac{3}{2} )(T-2Y)]
\,\rmd x'-Y \Big)
\\
=\tfrac{x}{x-1}{\mathring r},_{u }^{2}\Big(
\tfrac{4}{x(x-1)}
\int_{1}^x(x'-1)\Un{T}^\theta_\theta \,\rmd x'+
\tfrac{1}{x(x-1)}\int_{1}^xY\,\rmd x'
 -Y
\Big)
\\
\to \tfrac{x}{x-1}{\mathring r},_{u }^{2}\Big(
\tfrac{2(x-1)}{x}\Un{T}^\theta_\theta+
(\tfrac{1}{x}-1)Y +
\tfrac{x-1}{2x}Y'
\Big)
\\
\to
\tfrac{1}{2}{\mathring r},_{u }^{2}Y'\ogr{}{x=1}
\approx 10^{-4}m_0^{-4}{\mathring r},_{u }^{2}
\end{multline}
and, correspondingly,
\begin{equation}\label{eq:tau_2}
\tau_2(1) \approx 10^{-3}.
\end{equation}
  \rem To avoid confusion note that our coordinates $u$ and $v$ \emph{differ}
from those used \cite{ChrFu}. The latter --- let us denote them
$u_\text{CF}$ and $ v_\text{CF}$ ---
are related to the former by
\[
u_\text{CF}=-4m_0 \ln (-{u }),\qquad v_\text{CF}=4m_0 \ln {v }.
\]
\end{document}